# The Citation Impacts and Citation Environments of Chinese Journals in Mathematics

Ping Zhou[ab] and Loet Leydesdorff [b]

[a]Institute of Scientific and Technical Information of China,

15 Fuxing Road, Beijing, 100038, P. R. China;

zhoup@istic.ac.cn; pzhou@fmg.uva.nl.

[b]Amsterdam School of Communications Research (ASCoR),

University of Amsterdam, Kloveniersburgwal 48,

1012 CX Amsterdam, The Netherlands;

loet@leydesdorff.net; http://www.leydesdorff.net.

**Abstract**

Based on the citation data of journals covered by the *China Scientific and Technical Papers and Citations Database* (*CSTPCD*), we obtained aggregated journal-journal citation environments by applying routines developed specifically for this purpose. Local citation impact of journals is defined as the share of the total citations in a local citation environment, which is expressed as a ratio and can be visualized by the size of the nodes. The vertical size of the nodes varies proportionally to a journal's total citation share, while the horizontal size of the nodes is used to provide citation information after correction for the within-journal (self-) citations. In the "citing" environment, the equivalent of the local citation performance can also be considered as a citation activity index. Using the "citing" patterns as variables one is able to map how the relevant journal environments are perceived by the collective of authors of a journal, while the "cited" environment reflects the impact of journals in a local environment. In this study, we analyze citation impacts of three Chinese journals in mathematics and compare local citation impacts with impact factors. Local citation impacts reflect a journal's status and function better than (global) impact factors. We also found that authors in Chinese journals prefer international instead of domestic ones as sources for their citations.

## 1. Introduction

In recent years, the percentage world share of scientific publications of the P. R. of China has increased exponentially (Zhou & Leydesdorff, 2006). This trend is expected to continue since the Chinese government plans to invest more in science. A recently issued medium and long-term plan on scientific and technological development requires that Chinese central and local government investments in science and technology increase faster than China's economic growth rate (Jia, 2006). This strategy issued by the State Council of China will increase annual investment in R&D to 900 billion Yuan (US$112 billion) in 2020. This would boost the proportion of China's gross domestic product spent on R&D from today's 1.3% to 2.5%.

With this emphasis on S&T of the Chinese government, Chinese scientific journals will play increasing roles in scientific communication. However, as important communication channels for Chinese scientists, Chinese journals face fierce competition by their international counterparts. High-quality papers written by Chinese authors are usually submitted to international journals included in the *SCI* instead of domestic journals, because papers published by Chinese scientists in the *SCI* source journals can bring substantial rewards, such as professorships, research grants, and even housing (Jia, 2005). In order to improve the situation and help Chinese journals raise their qualities, the Chinese government has launched a campaign to boost Chinese journals. At the heart of the campaign is a new fund that will provide financial support to 300 to 500 scientific journals of the country. In addition, the government will encourage Chinese researchers to publish their results in domestic—rather than international—journals, and to place their results in free archives (Jia, 2005).

Since the campaign aims at supporting journals selectively, the policy question of how to evaluate journal quality becomes critical. Some may recommend selecting journals with higher impact factors in the same field and included in the *SCI*, the *China Scientific and Technical Papers and Citations Database* (*CSTPCD*), a database produced by the Institute of Scientific and Technical Information of China (ISTIC), or the *Chinese Science Citation Database* (*CSCD*), a database produced by Chinese Academy of Sciences (CAS) (Jin & Rousseau, 2004). Using such solutions to evaluate journal quality raises the following two problems: 1) journals with different citation patterns (i.e., disciplinary attribution) might be artificially classified as the same discipline; comparison among such journals is inappropriate; 2) Deliberate interventions by editors may weaken the objectivity of impact factors (Begley, 2006).

By using the routines introduced below, we hope to provide approaches for the Chinese government in the campaign of raising the quality of Chinese journals. The study focuses on two issues: disciplinary classification and local impact. We will compare the results achieved by using our routines with the results obtained with the currently available methods: inductive classifications and impact factors.

## 2. Materials and methods

### 2.1 Materials

In China, there are two databases engaged in the statistics of scientific publications. One is the *China Scientific and Technical Papers and Citations Database* (*CSTPCD*), and the other is the *Chinese Science Citation Database* (*CSCD*). In this study, we use the *CSTPCD* as the major data source. The *CSTPCD* was set up in 1988 analogously to the *Science Citation Index* (*SCI*) with the support of the Ministry of Science and Technology of China. The Institute of Scientific and Technical Information of China (ISTIC) has carried out and developed the project ever since, making the *CSTPCD* widely used by research institutions, scientific management organizations, and individual scientists to measure their research output (Wu *et al.*, 2004). When the database was first established in 1988, only 1,189 journals were included. Fifteen years later (2004), this number had increased to 1,608 journals. This corresponds to a share of approximately 36% of the scientific journals published in China (Ren, 2005).

Based on the *CSTPCD*, ISTIC publishes the *Chinese S&T Journal Citation Reports* (*CSTJCR*) annually, which is similar to the *Journal Citation Reports* (*JCR*) published by Institute of Science Information (ISI). We also use the *Science Citation Index* (*SCI*) to analyze the citation patterns of some Chinese journals in the international environment. We use data from 2003 and 2004, respectively. Routines that were recently further developed by us are used to extract matrices of journals that have either citing or cited relations. In order to visualize citation relations among journals, we use the visualization program Pajek which is available at http://vlado.fmf.uni-lj.si/pub/networks/pajek/.

## 2.2 Methods

Scientific journals cite each other forming an aggregated journal-journal citation matrix. These matrices contain interesting information like disciplinary similarities and journal hierarchies. One can normalize the matrices by using the cosine as the similarity measure (Salton & McGill, 1983). Salton's cosine is equivalent to the Pearson correlation coefficient (Jones & Furnas, 1987), but its non-parametric character has advantages in the case of sparse matrices (Ahlgren *et al.*, 2003). For the purpose of the visualization, it is convenient that the cosine provides us with positive values only, while Pearson correlation matrix also generates negative values. We will use the cosine between two vectors to measure similarities between the distributions of various journals included in a citation environment.

Through the routines one can obtain 'citing' and 'cited' matrices for each journal. In order to generate a citation matrix, we need first to select a journal—seed journal—to run the routines. A 'citing' matrix gathers journals that are cited by authors in the seed journal, while a 'cited' matrix is composed of journals whose authors cite the seed journal. Each journal citing or being cited to the extent of more than one percent is drawn into the matrix (Leydesdorff & Cozzens, 1993). The matrices are in ASCII format and can be read directly into Pajek. Cosine values below 0.2 were suppressed in order to enhance the interpretability of the visualizations.

In the Pajek picture, the width of the lines between nodes (i.e., journals) corresponds to the cosine value. If a line is thick, it indicates that the citation patterns of the journals at the vertices are similar; if a line is thin, the similarities of the citation patterns between the two journals are weak. Since journals in the same discipline can be expected to have similar citation patterns, journals linked by thick lines and in strong components can be classified into the same disciplinary category. If journals are not connected by lines, these journals can be expected to belong to different disciplines.

The size of the nodes reflects the percentage contribution to the citations in a matrix—citing or cited, respectively.[1] The grand sum of the citation matrix N (= $\sum c_{ij}$) is used as the basis for the normalization of the citation contributions. Each journal contributes with its margintotal $C_i$ (= $\sum c_i$) as a grandsum. In a local citation environment, a journal's contribution can be defined by this $C_i/N$ ratio. This ratio will be used for determining the size of nodes. By distinguishing between the vertical size and the horizontal one of a node, a second parameter can be used to indicate this percentage also after correction for within-journal (self) citations (Price, 1981; Noma, 1982). Thus, one is able to visualize how much a journal is dependent on an inner circle of authors citing one another by inspecting the shape of the nodes. Note that within-journal citations can be both self-citations of authors and citations among authors publishing in the same journal. In a citation picture, the shapes of nodes are consequently an ellipse in which the vertical and horizontal radii are different. The vertical radius of the ellipse indicates a journal's citation share in a specific environment, while the horizontal radius of a node informs us about a journal's share when within-journal citations are excluded.

[1] Because the ISI suppresses the single relations by summing them under the category 'all others', the citation matrix includes all citations above one. However, all citations, including single citations, are included in citation matrices from the *CSTPCD*.

In a 'citing' environment, the vertical radius represents the collective behaviour of authors in a journal in terms of percentage share of references in a local environment, while the horizontal radius represents a citation activity which maps whether authors in a journal are active in referring to other journals in the same environment. The larger a journal's horizontal radius, the more active authors in a journal would be in citing other journals.

In a 'cited' environment, the vertical radius represents a journal's citation impact, while the horizontal radius represents a journal's citation impact to other journals in a local citation environment after correction for within self-citations. A larger horizontal radius means a stronger local citation impact. The difference between the vertical and horizontal radii of the ellipse indicates the extent of a journal's within-journal citation. If the difference is large, the ellipse will be narrow. This indicates that the journal has a relatively large share of within-journal citations. However, if the difference is zero, the ellipse becomes a circle. This implies that the journal has no within-journal citations.

The narrower an ellipse, the more severe the effect of within-journal citations would be. When an ellipse becomes a vertical line in a reference environment, authors in a journal have little interest in referring to other journals in this environment. If such a degenerated ellipse appears in a cited environment, a journal has little citation impact to other journals in this environment.

The visualizations are based on the algorithm of Kamada and Kawai (1989) as it is available in Pajek. This algorithm represents the network as a system of springs with relaxed lengths proportional to the edge length. Nodes are iteratively repositioned to minimize the overall 'energy' of the spring system using a steepest descent procedure. (The procedure is analogous to some forms of non-metric multi-dimensional scaling.) A disadvantage of this model is that unconnected nodes may remain randomly positioned across the visualization. Since we use the symmetrical cosine-matrices as input for the visualizations, the graphs are not directed.

Using the routines mentioned above, all journals covered by the *CSTPCD* in 2003 and 2004 were mapped in terms of the cosines among the vectors of the journals in the environments of each seed journal. These matrices are available online at http://www.leydesdorff.net/. The relevant environment for each journal was determined by including all journals which cite or are cited by the journal under study to the extent of one percent of its citation rate in the respective dimension (He & Pao, 1986; Leydesdorff, 1986). This generates sets of the order of 10-50 journals, but we shall see that in the case of highly skewed distributions (for example, only within-journal citations) this procedure does not work.

## 3. Results

Since China's performance is most pronounced in the field of mathematics (DICCAS, 2004), journals in this field can be expected to provide us with the clearest pictures of how citation behavior may differ in domestic and international contexts. We selected a journal that has the highest impact factor in the category of mathematics classified by ISTIC in 2004, and compare the results with those for a journal with a much lower impact. We will explore the two issues of classification and impact through analyzing citation performance of the relevant citation environments of these journals in 2003 and 2004.

### 3.1 Citation patterns in the domestic environment

#### 3.1.1 Cited patterns of *JME* and *AMS-C*

According to the *CSTJCR* 2004, the *Journal of Mathematics Education* which is published in Chinese with the Chinese title *数学教育学报* (we will call it *JME* below for convenience) has the highest impact factor (0.84) among journals under the category of mathematics. Another journal, the *Acta Mathematica Sinica*, is also published in Chinese with the Chinese title of 数学学报 (we will call it *AMS-C* as abbreviation). The impact factor of *AMS-C* is 0.30, ranking the fifth in the same category of *JME*. The titles indicate already that the two journals may have different functions for the development of the discipline.

*a. The cited pattern of* *<u>JME</u>*

Figures 1 and 2 provide visualizations of the citation impact environments of *JME* in 2003 and 2004, respectively. In Figure 1, five journals cited *JME* but these journals' citation patterns are so different that there are no lines drawn between the vertices. This means that the citation patterns are not correlated. The ellipse of *JME* is very narrow, indicating that the journal has very heavy within-journal citations. In fact, the five journals in Figure 2 provided 259 references to *JME*, among which 239 were from *JME* itself. This corresponds with a share of 92%.

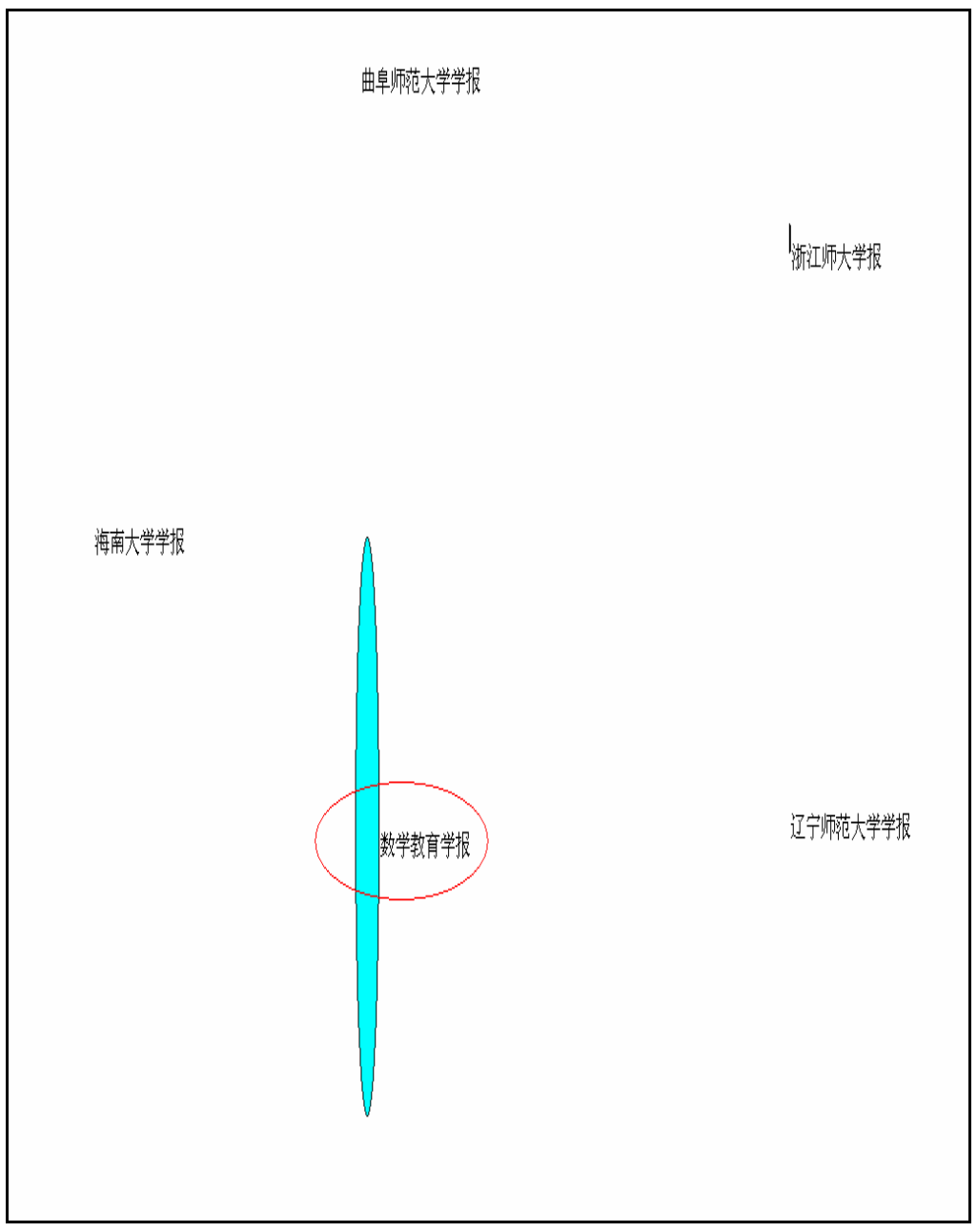

**Figure 1.** Citation impact environment of *Journal of Mathematics Education* (*数学教育学报*) in 2003 (*CSTPCD*, threshold = 1%; cosine ≥ 0.2).

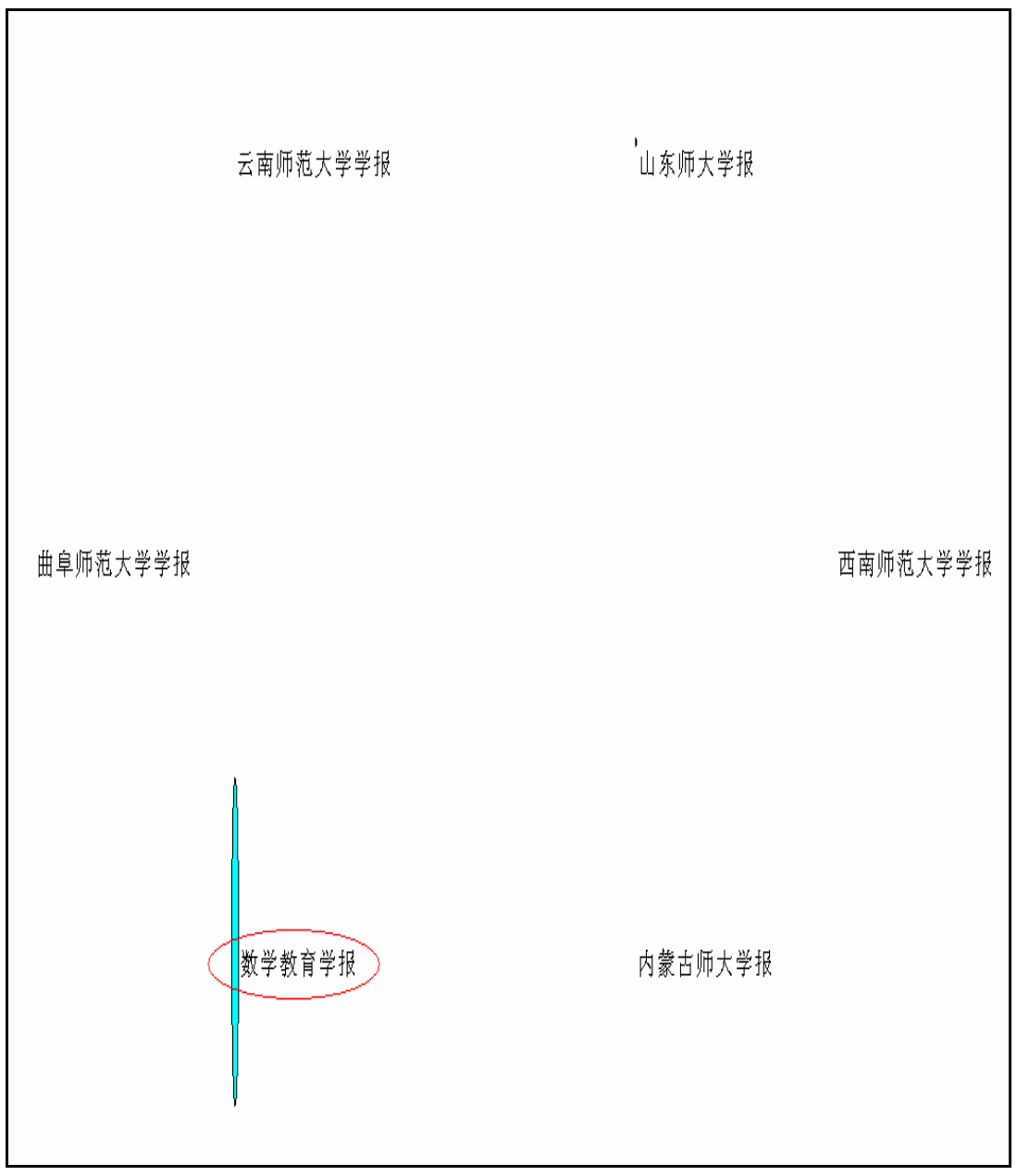

**Figure 2.** Citation impact environment of *Journal of Mathematics Education* (*数学教育学报*) in 2004 (*CSTPCD*, threshold = 1%; cosine ≥ 0.2).

In 2004, one more journal is included in the citation impact environment of *JME*. The ellipse of *JME* has become even narrower than in 2003, which means that the share of within-journal citations increased. In 2004, *JME* was cited to a total of 284 times among which 273 (96%) were provided by papers in *JME* itself.

The two figures show convincingly that most of citations to *JME* are from the journal itself, which implies that authors in the journal are not active in citing other journals. The other journals in its citation impact environment are university journals. Most of the university journals are from normal universities. The impact factors of these journals are very low, with ranks ranging from 58$^{th}$ to 84$^{th}$ among the 84 journals in the category of mathematics. University journals are multidisciplinary, publishing articles in various fields (Jin & Leydesdorff, 2005). Articles published in university journals and citing *JME* are in the field of mathematics. Thus we may conclude that the citation impact of *JME* is mainly within the journal, with a weak impact to normal university journals that publish articles in mathematics. This accords with the journal's title which emphasizes mathematical education.

*b. The cited pattern of* *<u>AMS-C</u>*

Figure 3 provides a visualization for the citation impact environment of *AMS-C* in 2003. Twenty-eight journals contribute to the total citation of *AMS-C* in this year for at least one percent.

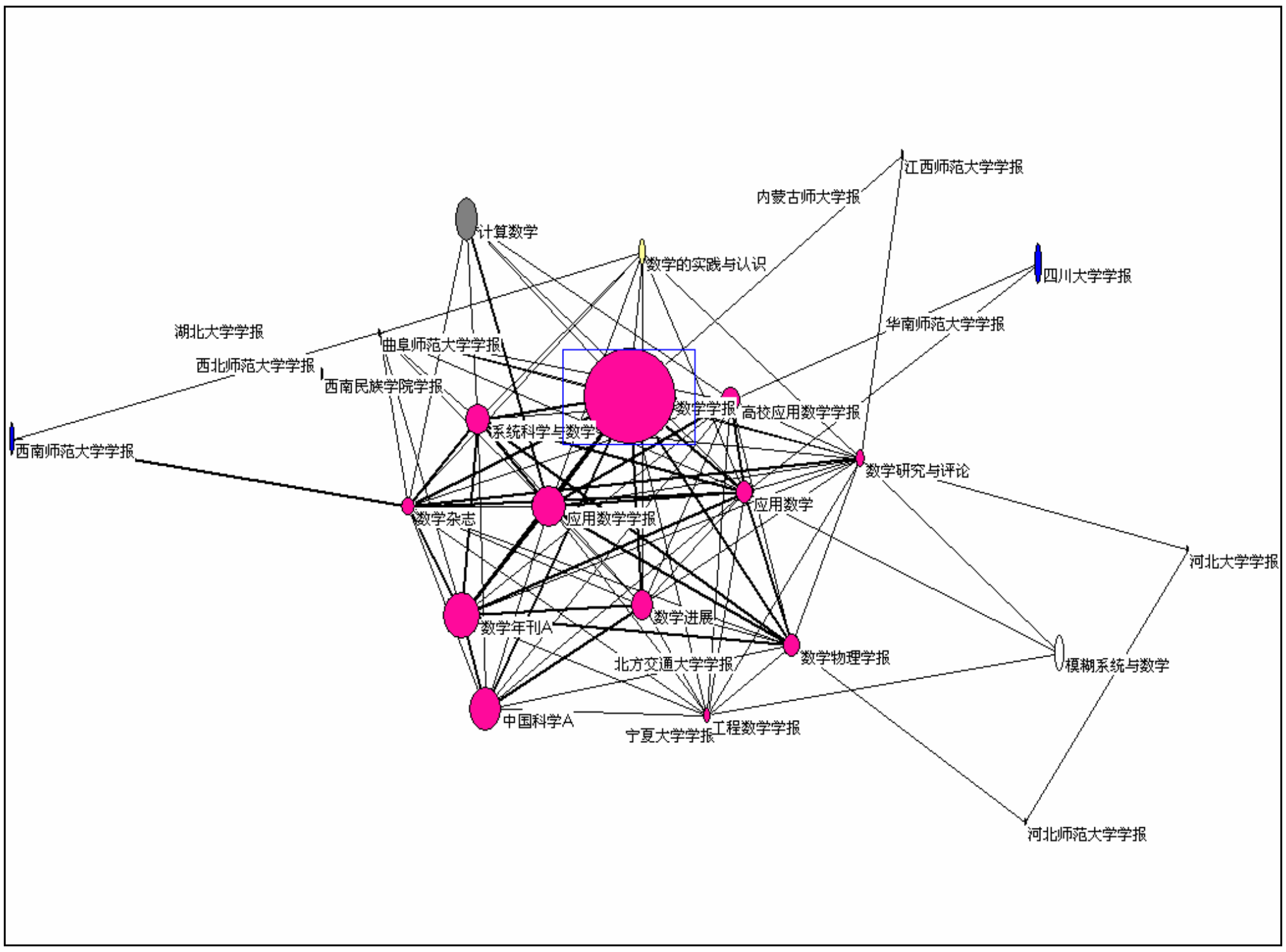

**Figure 3.** Citation impact environment of *Acta Mathematica Sinica* (*数学学报*) in 2003 (*CSTPCD*, threshold = 1%; cosine ≥ 0.2).

Journals in the center of the figure are linked by lines that are thicker than those among other journals. This means that the citation patterns of the journals in the center are similar. Therefore, these journals can be classified as belonging to the same discipline, i.e., mathematics in this case. In fact, all these journals have ‘mathematics’ in their titles. Journals in the periphery are university journals.

In Figure 3, the node that visualized *AMS-C* is a full circle, indicating that the within-journal citations of *AMS-C* in 2003 are zero. In addition, the node size of *AMS-C* is the largest in the figure, indicating that this journal has the highest local citation impact in its citation impact environment. The citation impact of *AMS-C* is distributed among mathematical journals and university journals.

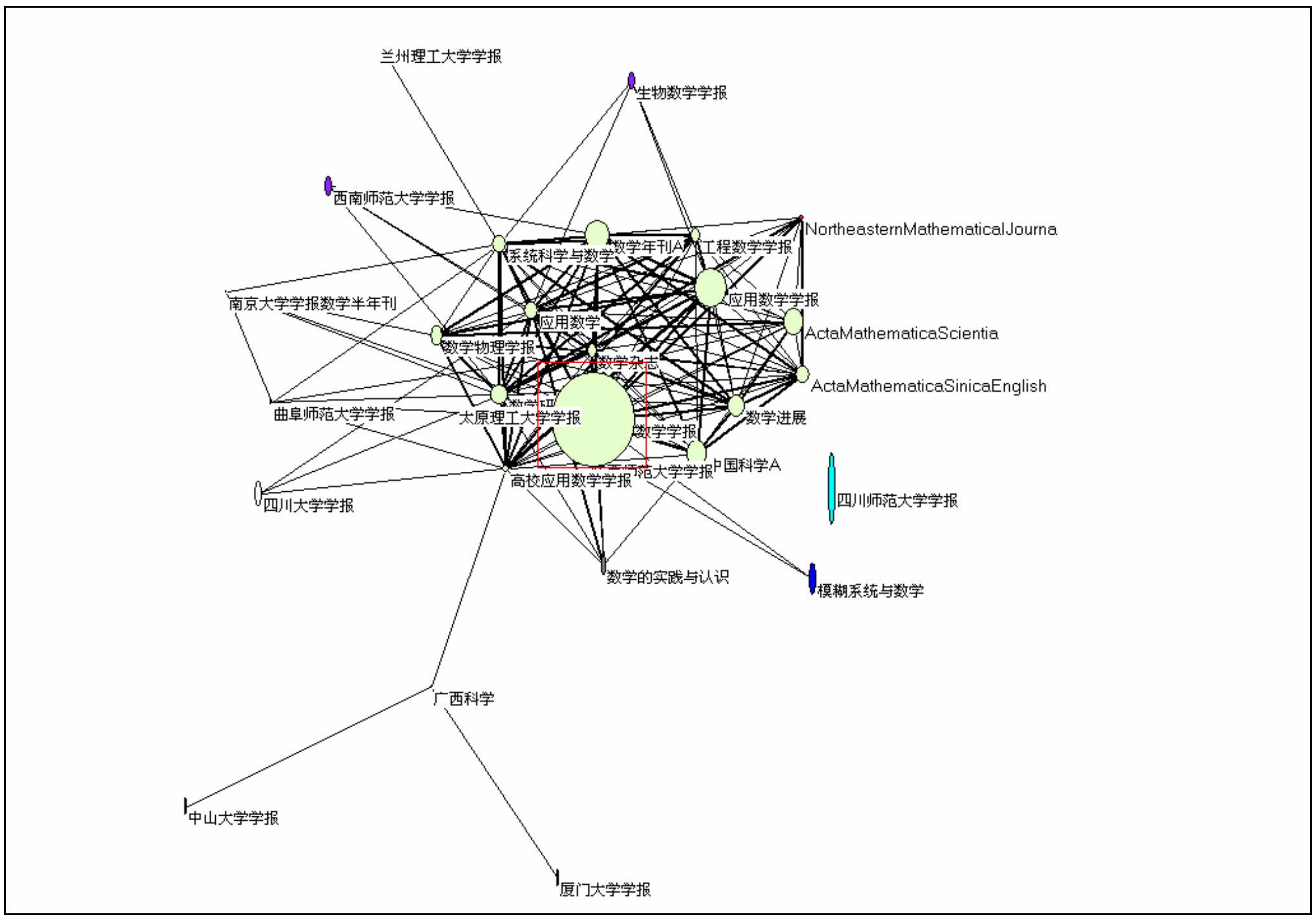

**Figure 4.** Cited environment of *Acta Mathematica Sinica* (数学学报) in 2004

(*CSTPCD*, threshold = 1%; cosine ≥ 0.2).

The citation impact of the journal of *AMS-C* changed slightly in 2004 (Figure 4). In addition to one more journals joining the citation impact environment, the node of *AMS-C* is no longer a full circle: the difference between the horizontal and vertical radii is more pronounced than in Figure 3. This indicates that there are some within-journal citations in 2004. Nonetheless, *AMS-C* still had the highest citation impact in the citation impact environment of *AMS-C*. The impact of *AMS-C* in terms of fields is similar to 2003. Journals linked by thick lines and gathered in the core are in mathematics, while university journals are distributed in the periphery.

Both *JME* and *AMS-C* are classified by the ISTIC as journals in mathematics. If we use impact factors as indicators for the comparison, *JME*, with an impact factor of 0.84, would rank higher than *AMS-C* which had an impact factor of 0.30. However, when we analyze their cited performances in relevant local environments, the impact of *AMS-C* is considerably higher than *JME,* for the following two reasons:

1) There are many more journals citing *AMS-C* than citing *JME*. In other words, *AMS-C* is more widely recognized among authors in other journals;

2) *AMS-C* influences more journals and more fields than *JME* does. The major impact of *AMS-C* is in mathematics journals, in addition to university journals.

From Figures 1, 2, 3 and 4 one can find interesting information: *JME* and *AMS-C* did not appear in each other's citation environments. As the journal that had the highest impact factor in the field of mathematics, *JME* had no citation impact to *AMS-C* whose impact factor ranked lower (the fifth in the same category of mathematics). Additionally, *JME* has a weak citation impact on normal university journals.

### 3.1.2 Citing patterns of *JME* and *AMS-C*

*a. Citing patterns of* *<u>JME</u>*

When the threshold is set at one percent, no other journals covered by *CSTPCD* are included in the reference environment of *JME* in either 2003 or 2004, except *JME* itself. This indicates three possibilities: 1) most of the references of *JME* were provided to itself; 2) journals cited by *JME* were not covered by the *CSTPCD*; and 3) combination of possibilities 1) and 2). The cited pattern of *JME* showed already that this journal has very heavy within-journal citation rates (Figures 1 and 2).

Upon closer inspection of the citation patterns, we found that *JME* cited only two journals covered by the *CSTPCD*. These two journals are *JME* itself and the *Journal of Liaoning Normal University*. Of the 274 references provided by *JME* to these two journals, 273 were within-journal citations. Only one reference was made to the *Journal of Liaoning Normal University*. Since the share of the *Journal of Liaoning*

*Normal University* in the total number of references of *JME* is less than one percent, the *Journal of Liaoning Normal University* does not appear in the reference environment of *JME*. Authors in *JME* provided 685 references in 2004.

In addition to the 273 references to the journal itself and one reference to another Chinese journals, the other 411 references were given to articles in international journals. In other words, the percentage of references of *JME* provided to international journals and *JME* itself journals are 60% and 40%, respectively. The citation impact of international journals on articles in *JME* is much larger than the impact of domestic journals.

In 2003, *JME* provided 555 references to journals, of which 245 were to domestic journals and 310 were to international journals. Among the 245 references, only 6 were given to five other domestic journals. The share of within-journal citations of *JME* in the domestic environment was 98%. International journals and *JME* itself accounted for 56% and 43% of the total references in articles published in *JME* during 2003.

In summary, authors in *JME* have little interests in referring other journals in the domestic environment except *JME* itself. This journal has very heavy within-journal citations. However, international journals are favorite among authors in *JME* in terms of making references.

*b. Citing patterns of AMS-C*

In addition to *AMS-C* itself, only two other Chinese journals covered by the *CSTPCD*, but published in English, were included in the reference environment of *AMS-C* in 2003 (Figure 5).

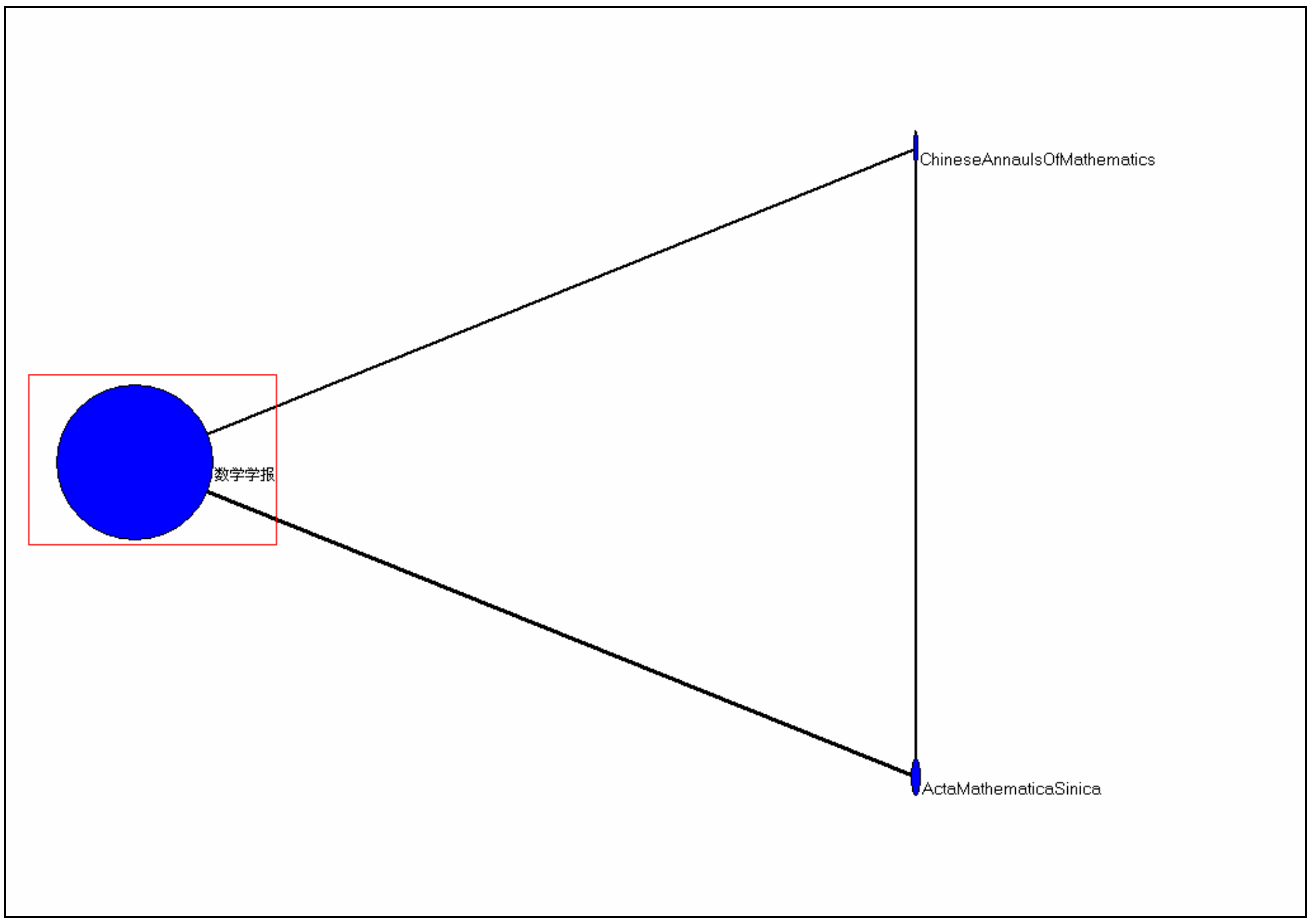

**Figure 5.** Reference environment of the *Acta Mathematica Sinica* (*数学学报*) in the *CSTPCD* in 2003 (threshold = 1%; cosine ≥ 0.2).

The English edition of *Acta Mathematica Sinica* is published by the same editorial board as *Acta Mathematica Sinica* (*数学学报*), but the articles are different. We will abbreviate the English journal of *Acta Mathematica Sinica* as *AMS-E*.[2] In Figure 5, the ellipse of *AMS-C* is almost a perfect circle and this vertex has the largest radius in

[2] In China, some journals are published in two languages, Chinese and English, but the English version is just a translation of the Chinese one.

the graph. This result teaches us four things:

1) The lines among the three journals in Figure 5 are thick, indicating that these journals have very similar patterns in providing references. Citation patterns of *AMS-C* and *AMS-E* are the most similar, since the line between these two journals is the thickest;

2) There are virtually no within-journal citations in *AMS-C* since the node shape of *AMS-C* is a circle;

3) In its local reference environment, *AMS-C* is the most active journal in terms of making references to other journals;

4) Both *Acta Mathematica Sinica* and *Chinese Annals of Mathematics Series B* have large shares of within-journal citations in this environment since the two ellipse are very narrow.

In order to check these conclusions, we ran the routines also using 2004 data. While we found that *AMS-C* did not provide any references to itself in 2003, some changes happened in the citing pattern of *AMS-C* in 2004. In this year, authors in *AMS-C* provided 112 references to journals covered by the *CSTPCD*, among which 49 were within-journal citations. The total number of references of *AMS-C*, however, was 1,209 in 2004. Journals covered by the *CSTPCD* just took a share of 9% of the references of *AMS-C*, while international journals accounted for the remaining 91% of the references.

In the domestic science community, the citing activity of the two journals show that *JME* mostly cited itself, while *AMS-C* is active in providing references to other journals, especially in 2003 when the journal did not provide a single reference to itself. *AMS-C* changed its citing pattern in 2004 a little bit by referring more often to articles published within it. However, within-journal citations do not constitute a large share of the total references in *AMS-C*. This share is just 9%. Authors in both *JME* and *AMS-C* prefer international journals in terms of making references. Over half of the references in *JME* are to international journals while this number is even higher for *AMS-C* (91% in 2004).

The citing patterns of the two journals show that international journals have a lager influence on Chinese journals than domestic ones. In the above analysis, international journals took 91% of the total references of *AMS-C* in 2004, while *JME* provided 40% of its total references to its international counterparts in that same year. Authors in *AMS-C* prefer citing internationally even more than authors in *JME*.

### 3.2 Comparison of citation patterns in domestic and international environments

In order to make the comparison more complete, we also selected *Acta Mathematica Sinica English Series* (*AMS-E*) as a seed journal. *AMS-E* is covered by both the *CSTPCD* and the *SCI*. Figures can thus be produced by using the journal as a seed in the two databases similarly.

#### 3.2.1 Cited patterns in the two databases

In the domestic environment, *AMS-E* is not only visible among mathematic journals, but is also cited in articles which are published in university journals. In the domestic citation impact environment of *AMS-E*, *AMS-C* has the highest local citation impact since the node of *AMS-C* is the largest (Figure 6). In Figure 4 (above), the local impact of *AMS-C* was also the largest. With regards to *JME*, the journal that has the highest impact factor among mathematic journals in 2004, does not appear in either of the two figures. This means that *JME* cites neither *AMS-C* nor *AMS-E*.

In the international citation impact environment of *AMS-E*, *AMS-E* is integrated with its international counterparts. In other words, *AMS-E* has international visibility. Nevertheless, the nodes of Chinese journals in the international environment are invisible (Figure 7), which implies that the local international impact of Chinese journals is weak.

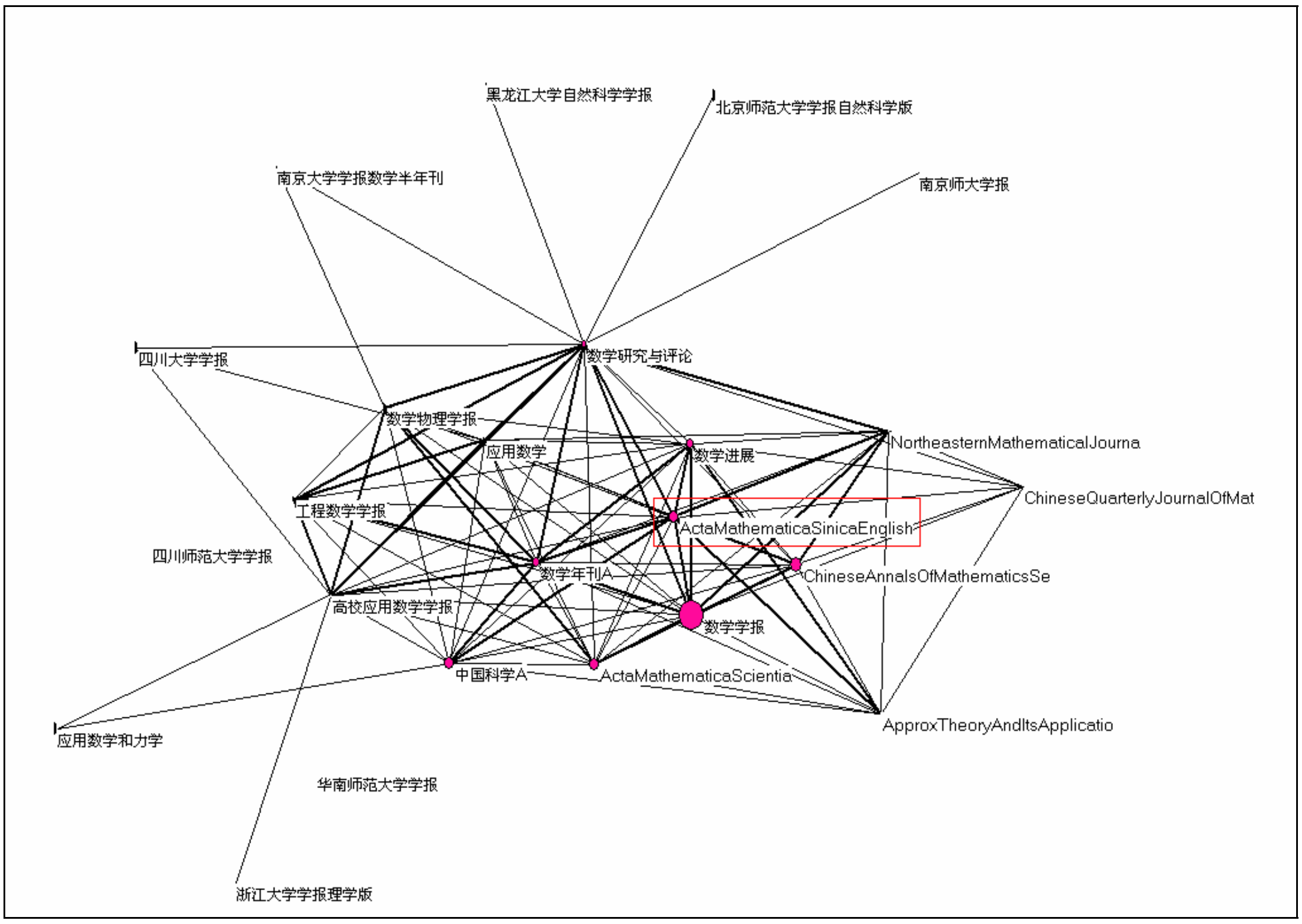

**Figure 6.** Citation impact environment of the *Acta Mathematica Sinica English Series* in 2004 (*CSTPCD*, threshold = 1%; cosine ≥ 0.2).

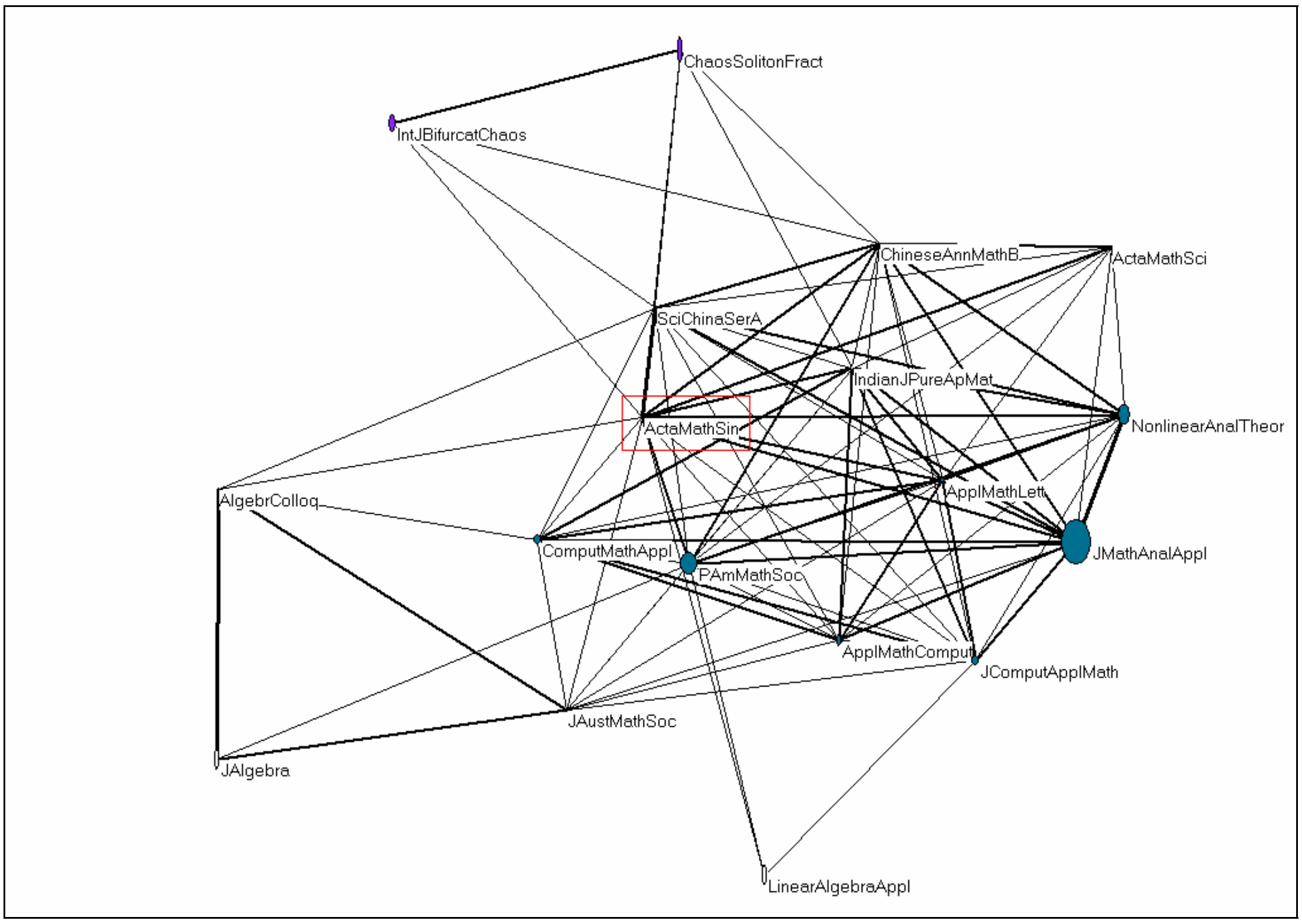

**Figure 7.** Citation impact environment of the *Acta Mathematica Sinica English Series* in 2004 (*SCI*, threshold = 1%; cosine ≥ 0.2).

### 3.2.2 Citing patterns in the two databases

Using the *CSTPCD*, only two journals are in the reference environment of *AMS-E* in 2004; that is: the journal itself and its counterpart in Chinese, *AMS-C* (Figure 8). Of the 1,273 references to journals, 81 were to domestic journals in the *CSTPCD* with a share of 6%. *AMS-E* did not cite *JME*, although *JME* had the highest impact factor in 2004 in the category of mathematics.

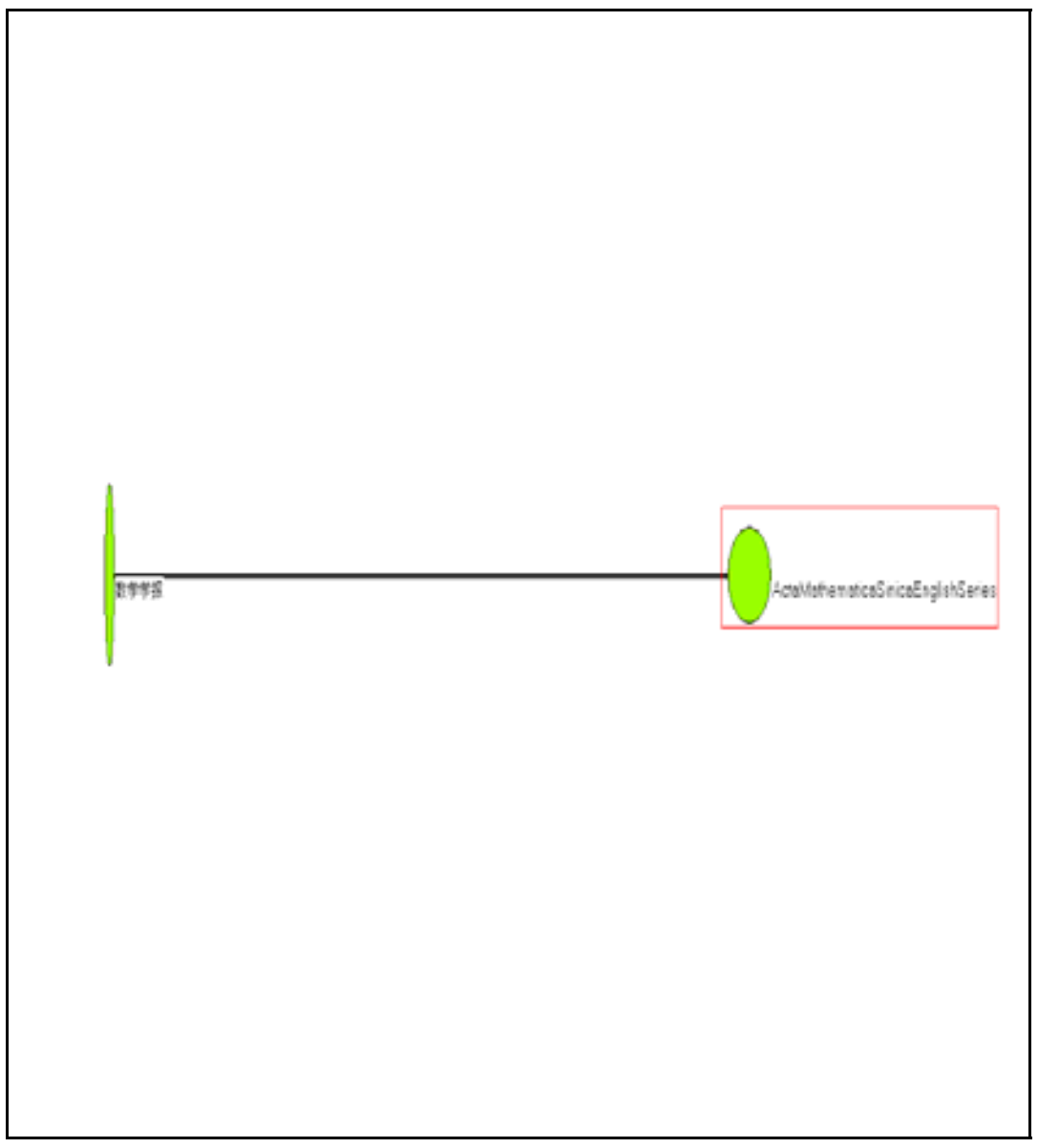

**Figure 8.** Domestic reference environment of the *Acta Mathematica Sinica English Series* in 2004 (*CSTPCD*, threshold = 1%; cosine ≥ 0.2).

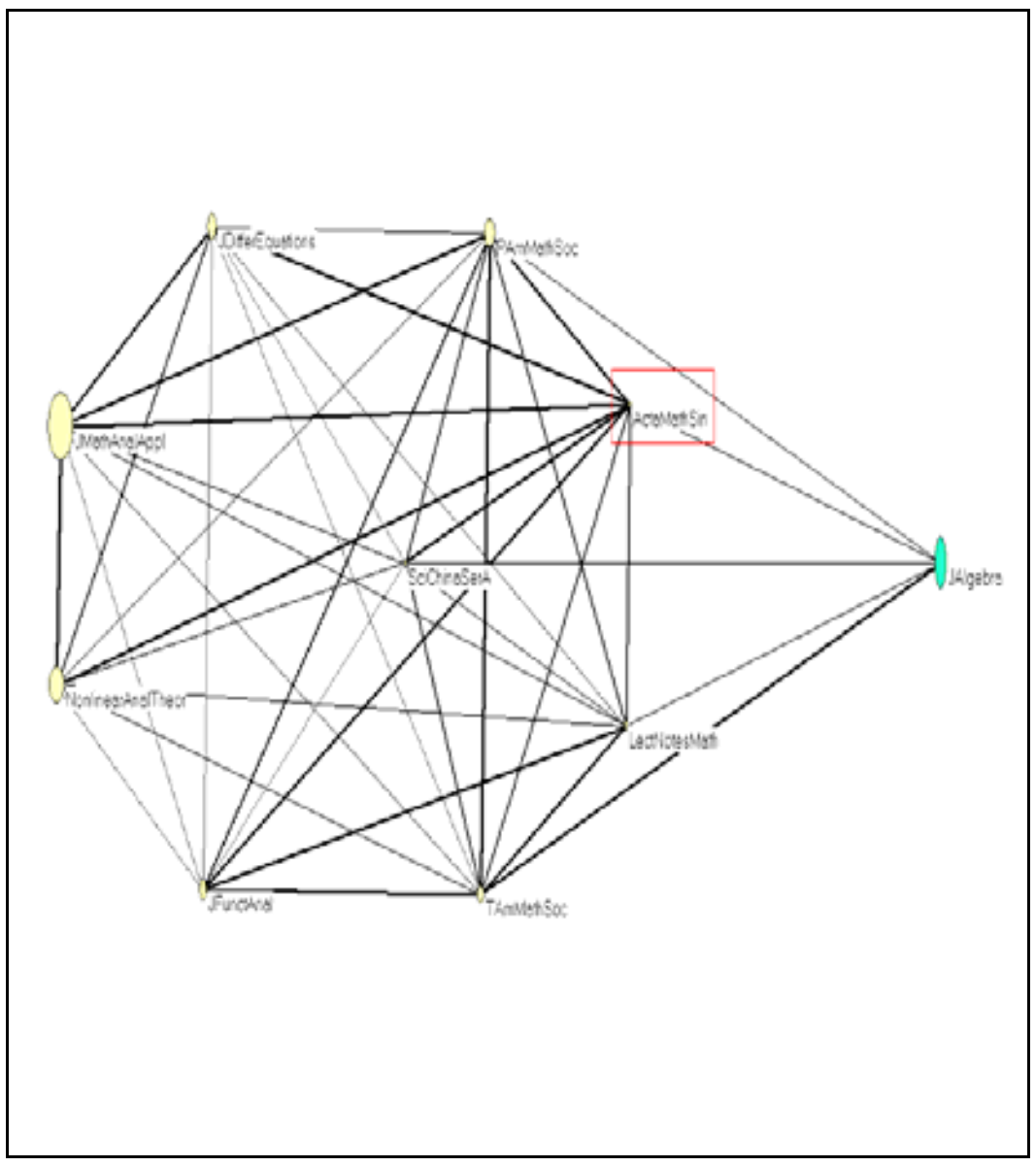

**Figure 9.** International reference environment of the *Acta Mathematica Sinica English Series* in 2004 (*SCI*, threshold = 1%; cosine ≥ 0.2).

Figure 9 provides the reference environment of the same journal (*AMS-E*) in the international environment based on the *SCI*. *AMS-E* cites more international journals than domestic ones in 2004. Except *AMS-E* and S*cience in China Series A* (a Chinese journal in mathematics), all the others are international journals. It is clear that authors in *AMS-E* favor international journals included in the *SCI* for making references. *AMS-E* gives 94% of its references to international journals. Compared to international journals, authors in Chinese journals make less references since the node sizes of the two Chinese journals in Figure 10 are very small while those of most international journals are relatively large.

In other words, Chinese journals in mathematics have a citation impact on both domestic and international journals. In the domestic environment, the citation impact can be seen among domestic mathematics journals and university journals. The local citation impact of *AMS-C* is the highest. In the international environment, however, the impact of Chinese mathematics journals is lower than that of their international counterparts.

Chinese authors in mathematics favor international journals included in the *SCI* for making references. Compared to their international counterparts, Chinese journals provide less references. Although *JME* has the highest impact factor among domestic journals in mathematics, there is no mutual citation interests between *JME* and the two language versions of *AMS*. This may imply that the research interests of authors in *JME* are not relevant to papers in *AMS-C* and *AMS-E*, and *vice versa*. This might be because *JME* is a journal in mathematics education, while *AMS-C* and *AMS-E* focus on research in mathematics.

## 4. Conclusions and discussions

In this study, we analyzed citation patterns of three Chinese journals in mathematics. These are the *Journal of Mathematics Education* and the Chinese and English versions of *Acta Mathematica Sinica* (labeled as *AMS-C* and *AMS-E* respectively).

Although the *Journal of Mathematics Education* (*JME*) has the highest impact factor in 2004 in the category of mathematics in the *CSTJCR*, the citation impact of *JME* is mainly within the journal itself. Its weak impact on other journals is limited to university journals whose impact factors are low. As a journal with the highest impact factor in the category of mathematics, *JME* has no citation impact on the group of journals classified as mathematics journals. Furthermore, *JME* does not appear in either the reference environments or citation impact environments of the two language versions of *AMS,* and *vice versa*.

*Acta Mathematica Sinica* (*AMS-C*) has the highest local citation impact in the field of mathematics although its global impact factor is much smaller than that of *JME*. In addition to university journals, *AMS-C* is well recognized by Chinese authors in mathematics. In the domestic environment, the two language versions of *AMS*, especially *AMS-C*, have impact on other mathematic journals while *JME*'s impact is limited to itself with a weak impact on normal university journals. *AMS-E* has visibility in the international environment, but its citation impact is weak compared to its international counterparts.

Both *JME* and the two language versions of *AMS* favor international journals for making references. Although authors in *JME* prefer to cite articles in this journal itself, it gave 60% of its total references to international journals in 2004. International journals take an even higher share of references in *AMS-C* than in *JME*. This number was 91% in 2004.

Compared to *AMS-C*, the communication dimension of *JME* is much narrower. On the one hand, *AMS-C* is more active in providing references to other journals. On the other hand, *AMS-C* is more recognized by other journals including international ones. According to its title, *JME* is a journal in mathematics education. Our analysis also shows that *JME* does not have citation impact on mathematics journals but does have citation impact on university journals, especially journals of normal universities which mainly teach future teachers. Therefore, perhaps it is better to put *JME* into a category of education science. Unfortunately, there is no such category available in the *CSTJCR*.

As an important journal engaging in higher education of mathematics, *JME* mainly serves for knowledge transfer, while *AMS-C* plays more role in knowledge generation.

A journal which is important for knowledge transfer has a higher impact factor than that of a journal which is central for knowledge generation in the same field.

In the category of mathematics in the *CSTJCR*, *JME* is the unique non-university journal for education. This situation makes *JME* an important communication platform and, therefore, an important reference source for articles in mathematics education. In other words, papers in mathematics education swarm into *JME* for publication. In the meantime, papers in *JME* become major reference sources of papers published in the journal, causing heavy within-journal citations.

The above analysis shows that evaluating a journal's quality is not a simple task (Leydesdorff, 2006). In addition to global impact factors, other elements like local citation impact and the specific role of a journal should be considered. Peer review is another important approach for this purpose. Furthermore, it is inappropriate to compare journals serving different functions. *JME* has the highest impact factor, but its citation impact on other journals is very weak. Although the impact factor of *AMS-C* is less than that of *JME*, *AMS-C*'s citation impact on other scientific journals is the largest in its local citation impact environment.

The fact that Chinese authors favor international journals may indicate that they assess the quality of articles published in Chinese journals as less then those published in international journals. Hence, improving quality of Chinese papers and journals is still a hard task for relevant players including authors, the government, and editorial boards.

## Acknowledgement

We would like to thank the statistics team of the Institute of Scientific and Technical Information of China, especially Ma Zheng of the team, for providing us with relevant data.